\documentclass[a4paper,10pt]{article}

\title{Monopoles in  Superloop Space }
\author{ Mir Faizal$^1$ and Tsou Sheung Tsun$^2$ \\
$^1$Department of Physics and Astronomy, \\  University of Waterloo,   Waterloo,\\
Ontario N2L 3G1, Canada \\
$^2$Mathematical Institute, University of Oxford,\\ Andrew Wiles
Building, \\Radcliffe Observatory Quarter, Woodstock Road,
\\ Oxford
OX2 6GG, United Kingdom 
 }
 \date{}
\begin{document}

\maketitle

\begin{abstract}
In this paper,  we will analyse a four dimensional gauge theory with $\mathcal{N} =1$ supersymmetry 
in superloop space formalism. We will thus obtain an expression for the connection in the 
infinite-dimensional superloop space. We will then use this connection to obtain an expression for the curvature 
 of the infinite-dimensional superloop space. We will also show that this curvature is proportional to 
 the Bianchi identity in spacetime. Thus, in absence of a monopole this curvature will vanish. 
 However, it will not vanish if the superloop intersects the world-line of a monopole because 
 the Bianchi will not hold at that point. 
\end{abstract}
\section{Introduction}
 Electric--magnetic duality in electromagnetism is an important
physical concept which has helped us understand, both the role of this
symmetry and topological concepts inherent in field theories.  Since
Dirac \cite{dirac} we know that the existence of magnetic monopoles is
equivalent to (electric) charge quantization which in turn is
equivalent to the electromagnetic gauge group being compact (i.\ e.\
$U(1)$).  In view of the crucial role of non-abelian gauge theories
in modern-day particle physics, the
question of a non-abelian version of duality and monopoles is much
studied \cite{nonabelian}-\cite{nona}.  As it turns out, loop space is crucial to
the study of these questions.  In this paper, we attempt to put these
questions in the context of superloops, as there are indications that 
some of the
physical applications in ordinary loop space might be usefully applied in
supersymmetric theories in various contexts to be given in more details below.

Monopoles in ordinary gauge theories have been studied using Polyakov loops (Dirac phase factor), 
which are closed loops in spacetime in which no trace is taken over the gauge group indices (unlike the Wilsons loops)
\cite{p1}. 
 Polyakov loop space has been generalized to $\mathcal{N} =1$  superloop space in  three dimensions
 \cite{mir}. It is interesting to further generalize this  to $\mathcal{N} =1$ superloop space in  four dimensions,
 because then it can be used to study various important physical systems. 
The non-abelian generalization of the Hodge duality (which is exactly
the duality of abelian gauge theory in four dimensions)
can only be constructed in loop space \cite{pq1}-\cite{pq2}.
Moreover, it is proved that this generalized loop 
space duality does reduce to the Hodge duality in spacetime for an abelian gauge theory. Furthermore, 
the sources in the original  theory,
appear as monopoles in the dual theory, and   monopoles 
in the original theory, become  sources in the dual theory.
So, this loop space  duality can be used to construct a dual potential for the non-abelian gauge theories. 
It may be noted that this can not be done in three dimensions as the original Hodge duality does exist in three dimensions. 
Thus, we need a four dimensional superloop space to construct a dual potential for supersymmetric gauge theories. 
It may be noted that  confinement problem in non-Abelian
gauge theories has been discussed  using 't Hooft's  order-disorder parameters \cite{1t}. 
The explicit construction of   these order-disorder parameters requires
  this dual potential  constructed in the loop space  \cite{1p}. So, four dimensional superloop space can be used to 
  study these order-disorder parameters in supersymmetric gauge theories. 

Furthermore, this dual potential  
has also been used for constructing a 
 Dualized Standard Model \cite{d}-\cite{d0}. 
Supersymmetry is also an important ingredient in construction of
particle physics models beyond the Standard Model. It may be noted
that the 
CKM matrix is not 
an identity matrix even in the supersymmetric Standard Models  \cite{cmk}-\cite{qqqq}. 
The results obtained in this paper can be used to explain this fact by 
constructing a supersymmetric Dualized Standard Model. In that supersymmetric Dualized Standard Model, 
 there will  
exist a dual gauge symmetry apart from the usual gauge symmetry. Thus, in the 
 supersymmetric Dualized Standard Model fermion generations will be expected to appear as dual colors.
 This symmetry is expected to be broken with interesting physical implications. 
 Furthermore, with simple assumptions
about the dual hypercharges of the dual Higgs fields, the fermion mass
hierarchy and that the CKM matrix are expected to be just the 
identity at the tree level. However, the loop corrections are expected to  give  small but non-zero values to both to the lower generation
fermion masses and the off-diagonal CKM matrix elements in the supersymmetric Dualized standard model. 
These results have already been obtained for the conventional Standard
Model 
using the duality derived from the loop space formalism \cite{d1}-\cite{d2}.
It will be interesting to obtain similar results for supersymmetric
Standard Model using the present superloop formalism.

 In this paper, we will construct  four dimensional  $\mathcal{N} =1$ superloop space. 
 From a supersymmetric point of view, this theory will have the same 
amount of supersymmetry as  three dimensional theory with $\mathcal{N} =2$ supersymmetry. So, the results of this paper  can be used 
for analysing monopoles in three dimensional gauge theories, in $\mathcal{N} =2$ superspace. 
It may be noted that in certain cases, monopoles in the ABJM theory can enhancement its supersymmetry  
from $\mathcal{N} =6$ supersymmetry to $\mathcal{N} =8$ supersymmetry \cite{abjm}-\cite{abjm1}. 
This is important as the theory of M2-branes is expected to have full $\mathcal{N} =8$ supersymmetry. 
It is important to study the effect of 
non-abelian monopoles in the ABJM theory for this  supersymmetry enhancement. 
Furthermore, it is expected that the study of monopoles in the ABJM theory can help 
us understand the physics behind multiple M5-branes \cite{m5}-\cite{5m}. 
The results of this paper can find direct applications for such analysis.

\section{Superloop Variables}

In this section we will construct Polyakov loops for four dimensional Yang-Mills theory with 
$\mathcal{N} =1$ supersymmetry. 
Polyakov loops are holonomies of closed loops in spacetime, and they are thus 
  defined via parametrized loops in spacetime. However, they 
 are independent of the parametrization chosen. As (
unlike Wilsons loops ) no trace is taken over gauge group indices in a 
Polyakov loop, they 
  are   gauge group valued functionals in the infinite-dimensional loop space.
Polyakov loop are sometimes  called the Dirac phase factor in the physics
literature.
As Polyakov loops are gauge group valued functionals, only their logarithmic derivative can be defined. 
This logarithmic derivative is used to define a connection in this loop space. This connection is called the 
Polyakov variable and it measures the change in phase  as one moves from one point in the loop space 
to a neighboring point \cite{p2,p3}. Furthermore, this connection can be used to construct a 
curvature tensor. 
This curvature tensor in the infinite-dimensional loop space is proportional to the 
Bianchi identity in spacetime, and so, it vanishes  when ever  the Bianchi identity is satisfied \cite{p4}. As the Bianchi 
identity is not satisfied in presence of a monopole, so it does not vanish if a monopole is present. 
A similar construction has been done for three dimensional dimensional Yang-Mills theory with $\mathcal{N} =1$ supersymmetry \cite{mir}, 
and in this section, we will generalize those results to a four dimensional Yang-Mills theory with $\mathcal{N} =1$ 
supersymmetry. 

A four dimensional Yang-Mills theory with $\mathcal{N} =1$ supersymmetry 
can be  parameterized by anticommuting coordinates 
$\theta^a$ and $\theta^{\dot{a}}$ along with  
the usual  spacetime 
coordinates, $x_{a\dot{a}} = \sigma_{a\dot{a}}^\mu x_\mu $ \cite{1001}. Here 
$\sigma_{a\dot{a}}^\mu = (\sigma_{a\dot{a}}^0, - \sigma_{a\dot{a}}^i  )$, with $\sigma_{a\dot{a}}^0$ being a two dimensional 
identity matrix and $\sigma_{a\dot{a}}^i$ being complex matrices with eigenvalues $\pm 1$ called the 
  Pauli matrices. 
The generators of $\mathcal{N} =1$ supersymmetry satisfy $\{Q_a, Q_{\dot{a}}\} = 
-i \partial_{a\dot{a}}$, where  $\partial_{a\dot{a}} = \sigma_{a\dot{a}}^{\mu}\partial_\mu$. 
It is useful to define the derivatives $D_a = \partial_a + 
i \theta^a \partial_{a\dot{a}}$ and $D_{\dot{a}} = \partial_{\dot{a}} + 
i \theta^{\dot{a}}\partial_{a\dot{a}}$, which commute with these generators of $\mathcal{N} =1$
supersymmetry. 
They also satisfy, $\{D_a, D_{\dot{a}}\} = 
i \partial_{a\dot{a}}$. 

Usually, a vector superfield, $V $, is used for 
constructing a four dimensional non-abelian gauge theory with $\mathcal{N} =1$ supersymmetry. 
This vector superfield is  a matrix valued superfield, $V= V^\alpha T_\alpha$, where 
$[T_\alpha, T_\beta] = i f_{\alpha \beta}^\gamma T_\gamma$, and it transforms under gauge transformation as 
$
\exp (V) \to \exp (i \bar \Lambda )\exp (V) \exp (-i \Lambda)
$.
It is possible to use a gauge called the Wess-Zumino gauge, where 
$[V]_| = [D_a V]_| = [D^a D_a V]_|  = [D_{\dot{a}} V]_| = [D^{\dot{a}}
D_{\dot{a}} V] _| =0  $, where $'|'$ means that the quantity is evaluated at $\theta =0$. 
Now we can construct a covariant derivative, $\nabla_A$ from this vector superfield as 
\begin{eqnarray}
 \nabla_A &=& (-i \{\mathcal{D}_a , D_{\dot{a}}\}, 
\mathcal{D}_a , D_{\dot{a}}), 
 \nonumber \\ 
 \exp ( V ) \nabla_A \exp( -V ) &=& (-i \{ D_a, \mathcal{D}_{\dot{a}} \}, 
  D_a , \mathcal{D}_{\dot{a}} ), 
\end{eqnarray}
where $\mathcal{D}_a = \exp (-V) D_a \exp (V )$ and $\mathcal{D}_{\dot{a}} =
\exp (V) D_{\dot{a}} \exp(- V )$.  
The transformation of this covariant derivative can now be written as 
$
 \exp ( V) \nabla_A \exp ( -V)  \to  \exp (i \bar \Lambda) \exp (V) \nabla_A \exp (-V) \exp( -i \bar\Lambda), 
$ and $
 \nabla_A \to \exp( i \Lambda) \nabla_A \exp (-i\Lambda) $.
Furthermore, from  the explicit 
form of the covariant derivative, the following quantities also vanish, 
$ F_{a\dot{a}} = F_{ab} =  F _{\dot{a}\dot{b} } =0$, and the remaining 
field strengths  can be expressed in terms of a single spinor  valued field strength, 
$2W_a = i D^{\dot{a}} D_{\dot{a}} \exp (-V )D_a \exp (V)$, where
$W_{\dot{a}} = - \exp( -V ) W_a^{\dagger} \exp (V)$ and  $\nabla^a W_a = - \nabla^{\dot{a}} 
W_{\dot{a}}$. Now we  can view the derivatives 
$D_A = (\partial_{a\dot{a}}, D_a, D_{\dot{a}} )$ as a supervector and thus define a super one-form as 
$\Gamma = \xi^{a\dot{a}}\Gamma_{a\dot{a}} + \xi^a\Gamma_a + \xi^{\dot{a}} 
\Gamma_{\dot{a}}$, 
where $\Gamma_{a\dot{a}} = \Gamma_{a\dot{a}}^\alpha T_{\alpha}, \, \Gamma_a =  
\Gamma_a^\alpha T_{\alpha},  \, \Gamma_{\dot{a}} = \Gamma_{\dot{a}}^\alpha T_{\alpha}$, and 
$[T_{\alpha}, T_{\beta}] = i f_{\alpha\beta}^\gamma T_\gamma$. 
The covariant derivative can also be defined as  $\nabla_A = D_A - i \Gamma_A$, such that 
$
 [\nabla_A, \nabla_B \} = H_{AB}
$, 
where  $H_{AB}=  T^C_{AB}\nabla_C - i F_{AB} $. 
Now the Bianchi identity can be written as 
$
 [\nabla_{[A}, H_{BC)}\} =0
$. 
As we can express  the field strength $F_{AB}$ as 
$
 F_{AB} = D_{[A} \Gamma_{B\}} - i [\Gamma_A, \Gamma_B\} - T_{AB}^C \Gamma_C, 
$
so, we can write 
$
 F_{a\dot{a}} = D_{[a} \Gamma_{\dot{a}\}} 
 + D_{[\dot{a}} \Gamma_{{a}\}} - i [\Gamma_a, \Gamma_{\dot{a}}\} - i \Gamma_{a\dot{a}}.
$
Now if we impose the constraint $F_{a\dot{a}} =0$, we obtain, $ i \Gamma_{a\dot{a}} =
D_{[a} \Gamma_{\dot{a}\}} 
 + D_{[\dot{a}} \Gamma_{{a}\}} - i [\Gamma_a, \Gamma_{\dot{a}}\} $. 
Thus, by imposing the constraint,  $F_{a\dot{a}} =0$, we can express $\Gamma_{a\dot{a}}$ 
in terms of $\Gamma_a$ and $\Gamma_{\dot{a}}$.

Now  we will construct a loop space formalism for 
super-Yang-Mills theories in four dimensions. 
Thus, we first  parameterizing the superloop space by the coordinates $\xi^A(s)  
= (\xi^{a\dot{a}}(s), \xi^a(s), \xi^{\dot{a}}(s))$, 
\begin{equation}
 C : \{ \xi^A (s): s = 0 \to 2\pi, \, \, \xi^A (0) = \xi^A(2\pi)\},  
\end{equation}
where   $\xi^A (0) = \xi^A(2\pi)$ is a fixed point in the superloop space. 
We now define a superloop space variable as, 
\begin{eqnarray}
 \Phi [\xi] &=& P_s \exp i \left(\int^{2\pi}_0 \Gamma^{a\dot{a}} (\xi(s))  \frac{d \xi_{a\dot{a}}}{ds} +  \Gamma^a (\xi(s))  
\frac{d \xi_a}{ds}   + \Gamma^{\dot{a}}(\xi(s))   \frac{d \xi_{\dot{a}} }{ds} \right) \nonumber \\ &=& 
P_s \exp i \int^{2\pi}_0  \Gamma^A (\xi(s)) \frac{d \xi_A}{ds}. 
\end{eqnarray}
where $P_s$ denotes ordering in $s$ 
increasing from right to left. 
The derivative in $s$ is  taken from below. It may be noted that labeling the superloop variable by a fixed point is over complete since 
the superloop variable $ \Phi [\xi] $, depends only on $C$ and not on the manner in which it is parameterized. 
Thus, if we introduce another parameter  $s' = f(s)$ instead of $s$, it will only change the variable in the integration and not its value. 
Thus, by using this new parameter there will be no change in the value of $\Phi [\xi]$.  
It may be noted that 
$\Phi[\xi] $ is a scalar superfield from the supersymmetric point of view.
Thus, it is possible to project out various ordinary loop superfields from it. 

Now as $\Phi [\xi]$ is a gauge group valued functional, we can only define its logarithmic derivative
\begin{equation}
 F_A [\xi| s] = i \Phi^{-1}[\xi] \frac{\delta}{\delta \xi^A (s)}\Phi[\xi],  
\end{equation}
where  $ F_A [\xi| s] = (F_{a\dot{a}}[\xi| s],
 F_a[\xi| s]),  F_{\dot{a}}[\xi| s])$. This quantity is a supersymmetric generalization of the Polyakov variable. 
 This acts like a connection in the superloop space. To see that we will first  define a parallel transport 
 from a point $\xi(s_1)$ to a point $\xi (s_2)$ as  
\begin{eqnarray}
 \Phi [\xi: s_1, s_2 ]  &=& P_s \exp i \left( \int^{s_2}_{s_1} \Gamma^{a\dot{a}} (\xi(s))  \frac{d \xi_{a\dot{a}}}{ds} +  \Gamma^a (\xi(s))  
\frac{d \xi_a}{ds}  + \Gamma^{\dot{a}}(\xi(s)) \frac{d \xi_{\dot{a}} }{ds} \right) \nonumber \\ &=& 
P_s \exp i \int^{s_2}_{s_1} \Gamma^A (\xi(s)) \frac{d \xi_A}{ds}. 
\end{eqnarray}
Now we will parallel transport  from a fixed point  along a fixed path  to another point say $s$. 
After reaching $s$, we will take a  detour then turn 
back along the same path till we reach the  original point where we started from. Thus, the phase factor generated by 
going along the path from the original point to $s$ will be canceled by the phase factor generated by 
going from $s$ to the original point. So, there will be no contribution from this path. However, there will be a 
contribution generated by  the transport along the infinitesimal circuit at $s$. 
 This  contribution will be 
 proportional to 
$H^{AB} (s)$, 
\begin{equation}
 F^A [\xi|s] =   \Phi^{-1}[\xi: s,0] H^{AB} (\xi (s) )
 \Phi^{-1}[\xi: s,0]\frac{d \xi_B (s) }{d s}, \label{la}
\end{equation}
As $ F^A [\xi|s] $ represents the change in phase of $\Phi[\xi]$ as one moves from one point in the superloop space 
to a neighboring point, we can regard it as a connection in the superloop space.
It may be noted that even though   $ F^A [\xi|s]$ proportional to the field strength in spacetime, 
it can be viewed as a  connection in the superloop space.
Now we first define a covariant derivative in superloop space as
\begin{equation}
 \nabla_A (s) = \frac{\delta}{\delta \xi^A (s)} + i  F_A [\xi|s]. 
\end{equation}
Now the curvature $ -i G_{AB}[\xi, s_1, s_2]$ 
of the loop space can be defined by taking a commutator of these two covariant derivatives, 
$
[\nabla_A [\xi(s_1)], \nabla_B [\xi(s_2)]]
$, 
\begin{eqnarray}
 G_{AB}[\xi ( s_1, s_2)] &=& \frac{\delta}{\delta \xi^A (s_2) }F_B [\xi|s_1]
- \frac{\delta}{\delta \xi^B (s_1) }F_A [\xi|s_2] \nonumber \\&&
+i [F_A [\xi|s_1], F_B [\xi|s_2]].
\end{eqnarray}

\section{Monopoles}
In the previous section we constructed a curvature  for the infinite-dimensional superloop space. 
In this section, we will use this curvature to analyse monopoles in the superloop space. 
In fact, it  has been  demonstrated that for a three dimensional
Yang-Mills theory with $\mathcal{N} =1$
supersymmetry, the superloop space curvature vanished in absence of a monopole \cite{mir}. 
Now we have prove  a four dimensional generalization of that result and show that  a four dimensional Yang-Mills theory with 
$\mathcal{N} =1$ supersymmetry, the superloop space curvature will vanish in absence 
of a monopole. In order to calculate the curvature explicitly, we will calculated 
the functional derivative of $F^B [\xi | s_1]$ with respect to $\xi_A (s_2)$. 
This derivative can be calculated by taking two infinitesimal variation of $ \Phi^{-1}[\xi_2] \Phi[\xi_3] 
- \Phi^{-1}[\xi] \Phi[\xi_1]$, where 
$
  \xi_3^A (s) = (\xi_3^{a\dot{a}} (s),\xi_3^{a} (s), \xi_3^{\dot{a}}(s) ) = \xi_1^A (s) + \delta '\xi ^A(s),\, \, 
  \xi_2^A (s) = (\xi_2^{a\dot{a}} (s),\xi_2^{a} (s), \xi_2^{\dot{a}}(s)) = \xi^A (s) + \delta '\xi ^A(s), \, \,  
\xi_1^A (s) = (\xi_1^{a\dot{a}} (s),\xi_1^{a} (s), \xi_1^{\dot{a}}(s))  = \xi^A (s) 
  + \delta \xi^A(s) $, and $\delta \xi^A  = \Delta^B \delta^A_B \delta (s- s'), \, \,  
  \delta' \xi^A  = {\Delta '}^B \delta^A_B\delta (s- s')$. 
Now  we using parallel transport along these paths, we  can write 
\begin{equation}
  \Phi[\xi_1] = \Phi[\xi] - i \int ds \Phi( \xi: 2\pi, s )
  \mathcal{H} (\xi(s))  \Phi(\xi: s, 0), 
\end{equation}
where 
\begin{eqnarray}
  \mathcal{H} (\xi(s))  &=& 
  H^{a\dot{a} b \dot{b}} (\xi(s)) \frac{d \xi_{ b \dot{b} }(s)}{ds} 
\delta \xi_{a \dot{a}} (s)  +
  H^{a b \dot{b}} (\xi (s)) \frac{d \xi_{ b \dot{b} }(s)}{ds} 
\delta \xi_{  a} (s)
\nonumber \\&&  + 
  H^{\dot{a} b \dot{b}} (\xi(s)) \frac{d \xi_{ b \dot{b} }(s)}{ds} 
\delta \xi_{   \dot{a}} (s) + 
  H^{a\dot{a} b } (\xi (s)) \frac{d \xi_{   b }(s)}{ds} 
\delta \xi_{  a \dot{a}} (s)  \nonumber \\ && + 
  H^{a\dot{a}  \dot{b}} (\xi (s)) \frac{d \xi_{    \dot{b} }(s)}{ds} 
\delta \xi_{  a \dot{a}} (s) +
  H^{\dot{a}  \dot{b}} (\xi (s)) \frac{d \xi_{    \dot{b} }(s)}{ds} 
\delta \xi_{   \dot{a}} (s)
\nonumber \\&&  + 
  H^{a b } (\xi (s)) \frac{d \xi_{   b  }(s)}{ds} 
\delta \xi_{  a } (s) + 
  H^{a \dot{b}} (\xi (s)) \frac{d \xi_{    \dot{b} }(s)}{ds} 
\delta \xi_{  a } (s) \nonumber \\ && + 
  H^{\dot{a} b } (\xi (s)) \frac{d \xi_{   b  }(s)}{ds} 
\delta \xi_{   \dot{a}} (s). 
\end{eqnarray}
Similarly, we can also write
\begin{equation}
  \Phi[\xi_2] = \Phi[\xi] - i \int ds \Phi[ \xi: 2\pi, s ]
  \mathcal{H} (\xi'(s))  \Phi[\xi: s, 0], 
\end{equation}
where 
\begin{eqnarray}
  \mathcal{H} (\xi'(s))  &=& 
  H^{a\dot{a} b \dot{b}} (\xi(s)) \frac{d \xi_{ b \dot{b} }(s)}{ds} 
\delta '\xi _{a \dot{a}} (s)  +
  H^{a b \dot{b}} (\xi (s)) \frac{d \xi_{ b \dot{b} }(s)}{ds} 
\delta '\xi _{  a} (s)
\nonumber \\&&  + 
  H^{\dot{a} b \dot{b}} (\xi(s)) \frac{d \xi_{ b \dot{b} }(s)}{ds} 
\delta '\xi _{   \dot{a}} (s) + 
  H^{a\dot{a} b } (\xi (s)) \frac{d \xi_{   b }(s)}{ds} 
\delta '\xi _{  a \dot{a}} (s)  \nonumber \\ && + 
  H^{a\dot{a}  \dot{b}} (\xi (s)) \frac{d \xi_{    \dot{b} }(s)}{ds} 
\delta '\xi _{  a \dot{a}} (s) +
  H^{\dot{a}  \dot{b}} (\xi (s)) \frac{d \xi_{    \dot{b} }(s)}{ds} 
\delta '\xi _{   \dot{a}} (s)
\nonumber \\&&  + 
  H^{a b } (\xi (s)) \frac{d \xi_{   b  }(s)}{ds} 
\delta '\xi _{  a } (s) + 
  H^{a \dot{b}} (\xi (s)) \frac{d \xi_{    \dot{b} }(s)}{ds} 
\delta '\xi _{  a } (s) \nonumber \\ && + 
  H^{\dot{a} b } (\xi (s)) \frac{d \xi_{   b  }(s)}{ds} 
\delta '\xi _{   \dot{a}} (s). 
\end{eqnarray}
Finally, we have
\begin{equation}
  \Phi[\xi_3] =  \Phi[\xi_1] - i \int ds \Phi[ \xi_1: 2\pi, s ]
  \mathcal{H}(\xi_1(s)) \Phi[\xi_1: s, 0],
\end{equation}
where 
\begin{eqnarray}
  \mathcal{H} (\xi_1(s))  &=& 
  H^{a\dot{a} b \dot{b}} (\xi_1(s)) \frac{d \xi_{1 b \dot{b} }(s)}{ds} 
\delta '\xi _{1a \dot{a}} (s) +
  H^{a b \dot{b}} (\xi_1(s)) \frac{d \xi_{1 b \dot{b} }(s)}{ds} 
\delta '\xi _{1a} (s)
\nonumber \\&&  + 
  H^{\dot{a} b \dot{b}} (\xi_1(s)) \frac{d \xi_{1 b \dot{b} }(s)}{ds} 
\delta '\xi _{1 \dot{a}} (s) + 
  H^{a\dot{a} b } (\xi_1(s)) \frac{d \xi_{1 b }(s)}{ds} 
\delta '\xi _{1a \dot{a}} (s)  \nonumber \\ && + 
  H^{a\dot{a}  \dot{b}} (\xi_1(s)) \frac{d \xi_{1  \dot{b} }(s)}{ds} 
\delta '\xi _{1a \dot{a}} (s)  +
  H^{\dot{a}  \dot{b}} (\xi_1(s)) \frac{d \xi_{1  \dot{b} }(s)}{ds} 
\delta '\xi _{1 \dot{a}} (s) 
\nonumber \\&&  + 
  H^{a b } (\xi_1(s)) \frac{d \xi_{1 b  }(s)}{ds} 
\delta '\xi _{1a } (s)+ 
  H^{a \dot{b}} (\xi_1(s)) \frac{d \xi_{1  \dot{b} }(s)}{ds} 
\delta '\xi _{1a } (s)  \nonumber \\ && + 
  H^{\dot{a} b } (\xi_1(s)) \frac{d \xi_{1 b  }(s)}{ds} 
\delta '\xi _{1 \dot{a}} (s). 
\end{eqnarray}
Here $\Phi[\xi_1: s, 0]$ can be written as 
\begin{eqnarray}
\Phi[\xi_1: s, 0] =&=&  \Phi[\xi: s, 0] - i \int_0^s ds' \Phi[ \xi: s, s' ]
\mathcal{H} (\xi(s'))  \Phi[\xi: s', 0] \nonumber \\ &&  +
i \Gamma^{A}(\xi(s)) \Phi[\xi: s, 0]\delta \xi_A (s),
\end{eqnarray}
where 
\begin{eqnarray}
  \mathcal{H} (\xi(s'))  &=& 
  H^{a\dot{a} b \dot{b}} (\xi(s')) \frac{d \xi_{ b \dot{b} }(s')}{ds'} 
\delta \xi_{a \dot{a}} (s')  +
  H^{a b \dot{b}} (\xi (s')) \frac{d \xi_{ b \dot{b} }(s')}{ds'} 
\delta \xi_{  a} (s')
\nonumber \\&&  + 
  H^{\dot{a} b \dot{b}} (\xi(s')) \frac{d \xi_{ b \dot{b} }(s')}{ds'} 
\delta \xi_{   \dot{a}} (s') + 
  H^{a\dot{a} b } (\xi (s')) \frac{d \xi_{   b }(s')}{ds'} 
\delta \xi_{  a \dot{a}} (s')  \nonumber \\ && + 
  H^{a\dot{a}  \dot{b}} (\xi (s')) \frac{d \xi_{    \dot{b} }(s')}{ds'} 
\delta \xi_{  a \dot{a}} (s') +
  H^{\dot{a}  \dot{b}} (\xi (s')) \frac{d \xi_{    \dot{b} }(s')}{ds'} 
\delta \xi_{   \dot{a}} (s')
\nonumber \\&&  + 
  H^{a b } (\xi (s')) \frac{d \xi_{   b  }(s')}{ds'} 
\delta \xi_{  a } (s') + 
  H^{a \dot{b}} (\xi (s')) \frac{d \xi_{    \dot{b} }(s')}{ds'} 
\delta \xi_{  a } (s') \nonumber \\ && + 
  H^{\dot{a} b } (\xi (s')) \frac{d \xi_{   b  }(s')}{ds'} 
\delta \xi_{ \dot{a}} (s').
\end{eqnarray}
and a similar expression for  $\Phi[\xi: 2\pi, s]$.
Now collecting all the variations, we obtain the following expression, 
\begin{eqnarray}
 \frac{\delta }{\delta \xi_A (s_2)} F^B [\xi | s_1] &=& 
 \Phi^{-1}[\xi: s_1, 0]   \nabla^B H^{AC} (\xi (s_2)) \nonumber \\ && \times\frac{d\xi_C (s_1)}{d s_1}\Phi [\xi: s_1, 0] \delta (s_2 - s_1)\nonumber \\ && + 
 \Phi^{-1}[\xi: s_2, 0]   H_{AB} (\xi (s_2)) \Phi [\xi: s_2, 0]\nonumber \\ && \times \frac{d}{d s_1}\delta (s_2 - s_1)
 \nonumber \\ && +i [F^A [\xi|s_2], F^B [\xi|s_1]]\theta (s_1 - s_2). 
\end{eqnarray}
Similarly, we can write 
\begin{eqnarray}
 \frac{\delta }{\delta \xi_B (s_1)} F^A [\xi | s_2] &=& 
 \Phi^{-1}[\xi: s_2, 0]   \nabla^A H^{BC} (\xi (s_1)) \nonumber \\ && \times \frac{d\xi_C (s_2)}{d s_2}\Phi [\xi: s_2, 0] \delta (s_1 - s_2)
 \nonumber \\ && + 
 \Phi^{-1}[\xi: s_1, 0]   H_{BA} (\xi (s_1)) \Phi [\xi: s_1, 0]\nonumber \\ && \times \frac{d}{d s_2}\delta (s_1 - s_2)
 \nonumber \\ && +i [F^B [\xi|s_1], F^A [\xi|s_2]]\theta (s_2 - s_1). 
\end{eqnarray}
Now  obtain an expression for the 
superloop space curvature, 
\begin{eqnarray}
G_{AB}[\xi, s] &=& \frac{\delta}{\delta \xi^A (s_2) }F_B [\xi|s_1]
- \frac{\delta}{\delta \xi^B (s_1) }F_A [\xi|s_2] \nonumber \\&&
+i [F_A [\xi|s_1], F_B [\xi|s_2]]\nonumber \\ &=& 
 \Phi^{-1}[\xi: s,0] [\nabla_{[A}, H_{BC)}\} \Phi[\xi: s,0]\frac{d\xi^C (s)}{ds}. 
\end{eqnarray}

Thus, we observe that the superloop space curvature is proportional to the Bianchi identity in the spacetime. 
Now if the Bianchi identity are satisfied, $[\nabla_{[A}, H_{BC)}\} =0$, then 
the superloop space curvature will vanish, $ G_{AB}[\xi, s] = 0$. 
The only non-vanishing contribution to the superloop space curvature can come from the existence of a monopole. 
This is because the Bianchi identity is not satisfied in presence of a monopole. So, if  the world-line 
of a monopole intersects the superloop space, then the curvature of the superloop space will not vanish. 
In other words the non-vanishing of the superloop space curvature is an indicator for the existence 
of a monopole in spacetime.

\section{Conclusion}

In this paper we have analysed a four dimensional non-abelian gauge theory with $\mathcal{N} =1$ supersymmetry 
in superloop space formalism. Thus, we were able to obtain an expression for a connection in this superloop 
space. An expression for the curvature of this superloop space was constructed using this connection.  
It was also demonstrated that this curvature is proportional to the Bianchi identity. Thus, it vanished if the 
Bianchi identity was satisfied. As the Bianchi identity is not satisfied in presence of a monopole, 
 this curvature also did not vanish if the world-line of a monopole intersected the superloop space. 
 In this way, the non-vanishing of the curvature acted as an indicator for the existence of a monopole in 
 spacetime.

 The results of this paper can be used for constructing a dual potential in supersymmetric gauge theories. This dual potential 
 can be used for constructing a supersymmetric Dualized Standard Model. It can also be used for constructing 
 't Hooft's  order-disorder parameters in supersymmetric gauge theories. The results obtained in this paper can also find important 
 applications in M-theory. This is because they can be used for studying supersymmetry enhancement in the ABJM theory. 
 They can also be used for understanding the physics of multiple M5-branes. 
 It may be noted that D-branes in a graviphoton background \cite{y1}-\cite{y2}, break the 
  $\mathcal{N} =1$ supersymmetry
to $\mathcal{N} =1/2$ supersymmetry \cite{sy}-\cite{y}.  
It will be interesting to analyse monopoles in such a deformed field theory in four dimensions. 
In order to do so, we will have to construct a deformed version of the superloop space.  
This can possible be done by replacing all the products of superfields with 
superstar products. It will be interesting to analyse the properties of monopoles in this deformed 
superspace.

\end{document}